\documentclass[aps,pra,twocolumn,groupedaddress,amsmath]{revtex4-1}

\usepackage{graphicx}
\usepackage{enumitem} \usepackage{calc}
\usepackage{hyperref} \hypersetup{ colorlinks, allcolors=[rgb]{.18,.19,.57} }

 \def\ket#1{\mathinner{|{#1}\rangle}}
\def\transition#1#2{\ket{5S_{1/2},F=#1}\rightarrow\ket{5P_{3/2},F^\prime=#2}}
\def\lowerket#1{\ket{5S_{1/2},F=#1}} \def\upperket#1{\ket{5P_{3/2},F^\prime=#1}}

\renewcommand*{\Re}{\operatorname{Re}} \renewcommand*{\Im}{\operatorname{Im}}
\begin{document}


\title{Velocimetry of cold atoms by matterwave interferometry}


\author{Max Carey} \email[]{max.carey@soton.ac.uk}
\affiliation{School of Physics \& Astronomy, University of Southampton,
  Highfield, Southampton SO17 1BJ, United Kingdom}

\author{Jack Saywell}
\affiliation{School of Physics \& Astronomy, University of Southampton,
  Highfield, Southampton SO17 1BJ, United Kingdom}

\author{David Elcock}
\affiliation{School of Physics \& Astronomy, University of Southampton,
  Highfield, Southampton SO17 1BJ, United Kingdom}

\author{Mohammad Belal}
\affiliation{School of Physics \& Astronomy, University of Southampton,
  Highfield, Southampton SO17 1BJ, United Kingdom}

\author{Tim Freegarde}
\affiliation{School of Physics \& Astronomy, University of Southampton,
  Highfield, Southampton SO17 1BJ, United Kingdom}


\date{\today}

\begin{abstract}
  We present an elegant application of matterwave interferometry to the
  velocimetry of cold atoms whereby, in analogy to Fourier transform
  spectroscopy, the 1-D velocity distribution is manifest in the frequency
  domain of the interferometer output. By using stimulated Raman transitions
  between hyperfine ground states to perform a three-pulse interferometer
  sequence, we have measured the velocity distributions of clouds of
  freely-expanding $^{85}$Rb atoms with temperatures of 33~$\mu$K and 17~$\mu$K.
  Quadrature measurement of the interferometer output as a function of the
  temporal asymmetry yields velocity distributions with excellent fidelity. Our
  technique, which is particularly suited to ultracold samples, compares
  favourably with conventional Doppler and time-of-flight techniques, and
  reveals artefacts in standard Raman Doppler methods. The technique is related
  to, and provides a conceptual foundation of, interferometric matterwave
  accelerometry, gravimetry and rotation sensing.
\end{abstract}


\maketitle

\section{\label{sec:introduction}Introduction}

In the macroscopic world, we measure an object's temperature by bringing it into
thermal equilibrium with a small probe. When the object is a cloud of cold
atoms, the tiny thermal mass requires a thermodynamic probe on the atomic scale
\cite{Hohmann2016}; and inhomogeneous cooling, incomplete thermalization,
coupling to electronic energy levels, and non-equilibrium quantum thermodynamics
can mean that the temperature is poorly defined \cite{Millen2016}. It is
therefore common to characterize a cold atom sample by its velocity
distribution, from which the kinetic temperature may be found by subsequent
parameterization.

Popular methods of measuring velocity distributions, such as time-of-flight
imaging \cite{Lett1988}, Raman \cite{Kasevich1991,Hughes2017} and Bragg
\cite{Stenger1999a,Deh2009} Doppler spectroscopy, rely upon separate
interactions with small slices of the velocity distribution to build up a
complete measurement. Each involves the signal from only a small number of
atoms, and there are artefacts from the initial cloud size and
measurement-induced perturbation of the velocity distribution. Despite a variety
of enhancements \cite{Lu2011,Zhang2011,Pradhan2008}, precise interpretation of
the results requires care \cite{Hughes2017,Brzozowski2002}.

Here we present an alternative method of cold atom velocimetry by matterwave
interferometry, in which broadband interactions with laser pulses allow all the
atoms within the cloud to be interrogated simultaneously, permitting a
significant improvement in the measurement signal-to-noise ratio while
subjecting the sample to only the smallest perturbations and constraints. In
contrast to Doppler spectroscopy, the technique is particularly suited to the
lowest temperatures. While it can be considered a method of measuring the
temperature-dependent coherence length of the atomic wavepacket
\cite{Featonby1998, Saubamea1997, Marechal2000}, we show here that it allows the
velocity distribution to be measured in detail. It provides an elegant
conceptual underpinning of matterwave accelerometry, gravimetry and rotation
sensing, in which differential velocity measurements are performed by two
sequences in succession
\cite{Riedl2016,Peters2001,Gustavson2000,Kasevich1991a,Barrett2016}, and as a
result inspires adaptations that for example offer immunity to
mechanically-induced laser phase noise between the two measurements
\cite{DAmico2018}.
\section{Conceptual overview\label{sec:conceptual-overview}}

\begin{figure}[b]
  \centering
  \includegraphics{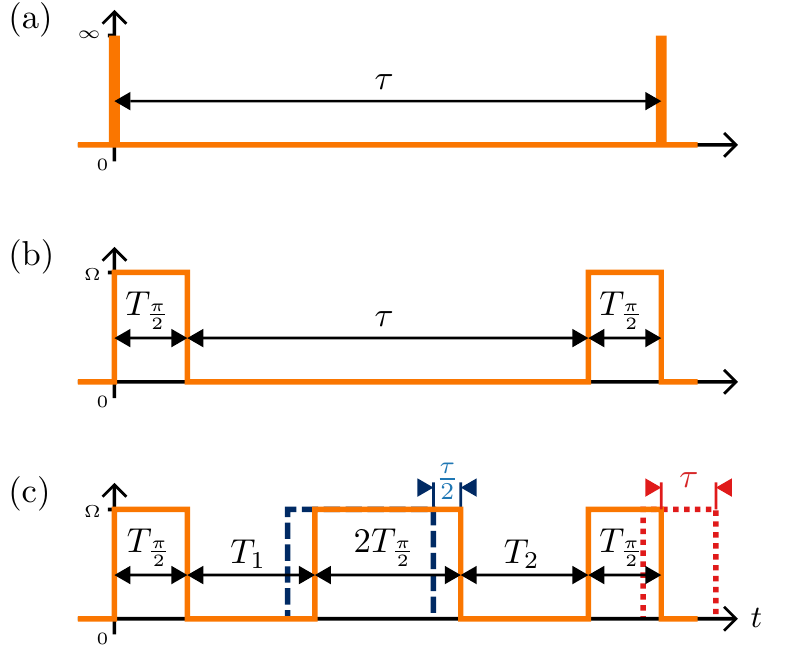}
  \caption{\label{fig:sequences}Temporal pulse profiles for the interferometer
    sequences considered here. a) Ramsey sequence with ideal, zero-length,
    pulses. b) Ramsey sequence with realistic, finite-length, pulses. c)
    Asymmetric Mach-Zehnder sequence with realistic pulses.}
\end{figure}

\begin{figure*}[!t]
  \includegraphics{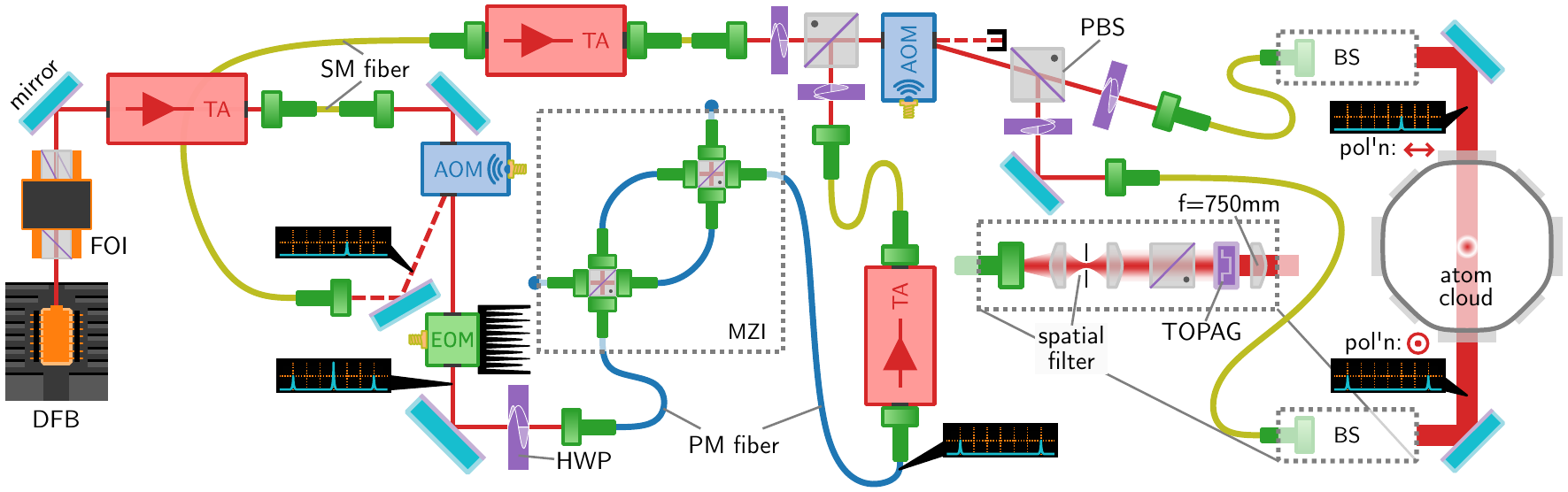}
  \caption{\label{fig:apparatus}Schematic for Raman interferometer: Distributed
    feedback laser diode (DFB), Faraday optical isolator (FOI), tapered
    amplifier (TA), acousto-optic modulator (AOM), electro-optic modulator
    (EOM), half waveplate (HWP), polarizing beam-splitter (PBS), Topag GTH-4-2.2
    refractive beam-shaper (TOPAG), beam-shaping optics (BS), single-mode (SM)
    and polarization-maintaining (PM) fibers, fiber Mach-Zehnder filter (MZI).
    MOT optics are not shown, and only one instance is labeled for some repeated
    symbols.}
\end{figure*}

Matterwave interferometry is performed by sequences of pulsed interactions,
resonant with the two-state quantum system $(\ket{1},\ket{2})$, that are
separated by periods of free evolution \cite{Borde1989a}. Prior to each
sequence, the atoms are pumped into one of the two states, $\ket{1}$. The first
pulse then places the atom into a quantum superposition of the two states, whose
phase is determined by that of the interaction field. The effect of subsequent
pulses then depends upon the phase difference accrued between the timebase of
the interaction field and the atomic state \cite{Gustavson2000}.

An example is the Ramsey interferometer \cite{Ysi1950}, consisting of two pulses
separated by a free-evolution period $\tau$ as depicted in Figure
\ref{fig:sequences}a. Each ``$\pi/2$-pulse'' lasts a quarter of the Rabi
oscillation period, and thus converts either of the two atomic states into an
equal superposition. In the case that $\tau=0$, the two pulses combine to form a
``$\pi$-pulse'', which transfers atoms from one state to the other.

For $\tau>0$, the first pulse leaves each atom in a superposition which then
evolves freely to accrue a relative phase $\Phi$ according to the atoms'
environment and trajectory. The phase determines the effect of the second pulse
of the sequence, and can thus be measured by monitoring the atomic state
populations once the sequence is complete.

If the interaction is with a pulsed laser beam, then atoms propagating with
velocity $v_z$ along the beam axis will experience a Doppler-shifted field which
gives the superposition phase $\Phi$ a dependence on the distance traveled
during the evolution time $\tau$. If there are no other phase contributions then

\begin{equation}
  \label{eq:phase shift}
  \Phi = \mathbf{k}\cdot\mathbf{v} \tau = k v_z \tau,
\end{equation}

\noindent where $k$ is the laser wavenumber \cite{Bongs2002,berman1996}. The
second pulse maps this phase onto atomic state probabilities so that, with ideal
pulses that perform this mapping exactly, the probability that a given atom is
in the second atomic state $\ket{2}$ is

\begin{equation}
  \label{eq:single atom output}
  |c_2|^2 = \frac{1}{2}[1 + \cos(k v_z \tau)],
\end{equation}

\noindent and thus the fraction of a statistical ensemble in state $\ket{2}$
will be

\begin{equation}
  \label{eq:quadratureoutput}
  \mathcal{S}(\tau) = \int\limits_{-\infty}^{\infty}P(v_z)\frac{1}{2}[1+\cos(k v_z \tau)] dv_z
\end{equation}

\noindent where $P(v_z)$ is the normalized distribution of velocity components
$v_z$ in the beam direction. Each velocity class $v_z$ hence contributes to the
interferometer signal a component which varies sinusoidally with $\tau$, with
frequency $k v_z$ and amplitude proportional to $P(v_z)$, akin to the
contributions from different wavelengths of light to the signal produced when
varying the arm length of a Michelson interferometer in Fourier transform
spectroscopy \cite{Michelson1891,bell1972introductory}. The velocity distribution is
thus mapped onto the frequency domain of the signal but, owing to the symmetry
of the cosine function, positive and negative velocities cannot be
distinguished.

If the laser phase is advanced by $\phi$ between the two interferometer
pulses, this phase is mapped onto the output signal such that

\begin{equation}
  \label{eq:inphaseoutput}
  \mathcal{S}(\phi, \tau) = \int\limits_{-\infty}^{\infty}P(v_z)\frac{1}{2}[1+\cos(k v_z \tau - \phi)] dv_z.
\end{equation}

\noindent The absolute value of the Fourier transform of the quantity

\begin{equation}
  \label{eq:combinedsignal}
  \mathcal{S}_I + i \mathcal{S}_Q \equiv \mathcal{S}(0, \tau) + i \mathcal{S}(\pi/2, \tau)
\end{equation}

is then proportional to the velocity distribution $P(k v_z)$, expressed as a
function of the frequency $k v_z$ \cite{Carey2017, Note1}.

\section{A Mach-Zehnder interferometer for atom
  velocimetry\label{sec:mach-zehnd-interf}}

The astute reader will notice that the Fourier transform of the signal defined
in Equation (\ref{eq:combinedsignal}) can only be measured experimentally for
positive pulse separations $\tau$, and that the signal is thus effectively
multiplied by a Heaviside step function. With ideal interferometer pulses that
perform perfect, instantaneous operations upon all atoms (Figure
\ref{fig:sequences}a) this would introduce an orthogonal component to the
Fourier transform that could be separated from the velocity information, but in
practice deconvolution becomes intractable because pulses of finite duration
(Figure \ref{fig:sequences}b) themselves exhibit Doppler sensitivity,
introducing a velocity-dependent amplitude and phase shift that we have explored
in more detail in \cite{Carey2017}.

In this work we therefore interleave a ``mirror'' $\pi$--pulse between the
``beam-splitters'' of our interferometer, as shown in Figure
\ref{fig:sequences}c. In its time-symmetrical form ($T_2=T_1$), atoms divide
their time equally between the two interferometer states, and the interferometer
forms a basic ``composite pulse'' \cite{Dunning2014} in which systematic
contributions to the phase accrued during the first evolution period $T_1$ are
reversed during the second period $T_2$. This Mach-Zehnder arrangement forms the
basis for atom interferometric inertial sensing, since steady velocity-dependent
phase shifts cancel and only the phase shifts due to changes in velocity remain
\cite{Kasevich1991a,Gustavson2000,Peters2001,Riedl2016,Barrett2016}.

When the interferometer is asymmetric, however, we retain the velocity
sensitivity according to the temporal asymmetry $\tau = T_2 - T_1$, which can be
varied continuously over both negative and positive values, while taking
advantage of partial cancellation of phase shifts accrued during the pulses
themselves \cite{Gillot2016}.

To maintain a constant atom cloud expansion in our experiments, we set the total
interferometer duration $T=T_1+T_2$ to a constant and vary $\tau$ between $-T$
and $T$. We show in Appendix \ref{sec:analyt-expr-interf} that some Doppler
sensitivity remains in a velocity-dependent modification of the fringe amplitude
that could be corrected for in subsequent analysis, and the introduction of
sub-harmonics with twice the fringe period that enhance the apparent probability
of lower atomic velocities. Provided the Doppler shift is no more that 0.4 of
the Rabi frequency, however, these effects are negligible.

\section{Experimental procedure\label{sec:experimental-setup}}

\begin{figure}[t]
  \includegraphics{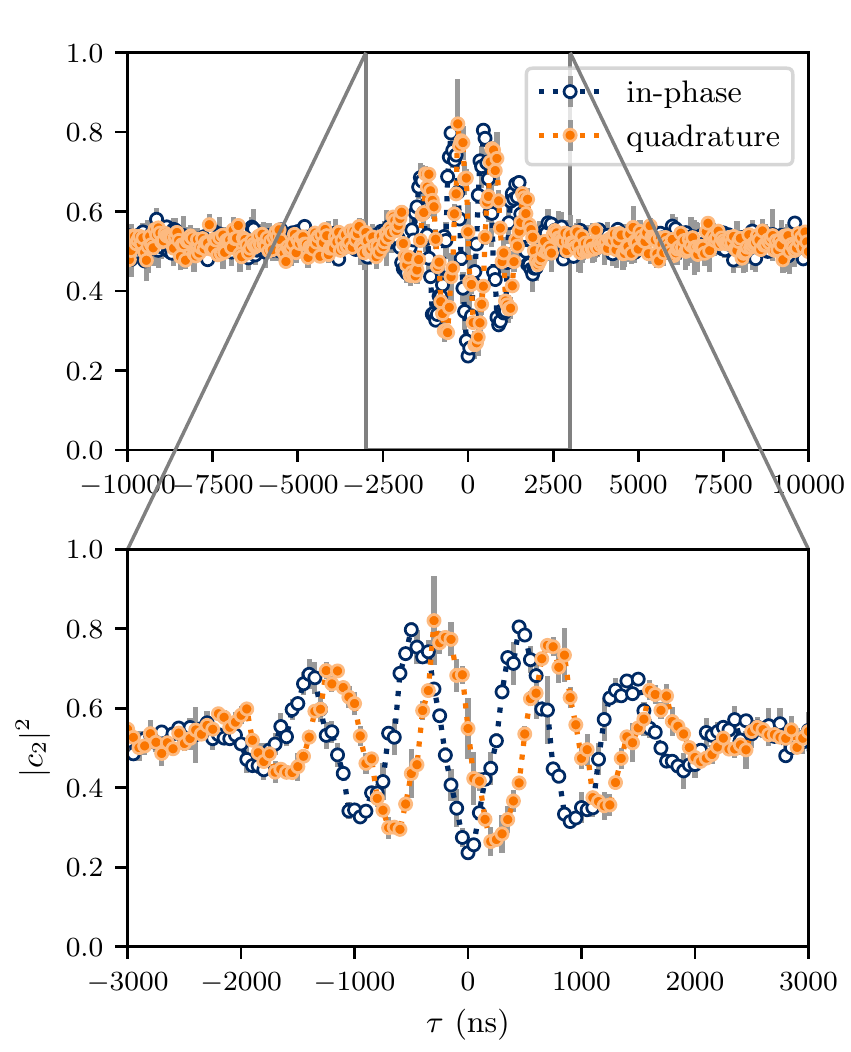}
  \caption{\label{fig:data1}Measurements of fractional population transfer
    $|c_2|^2$ as a function of temporal asymmetry $\tau$ for interferometers
    with (filled orange circles) and without (open blue circles) a $\pi/2$ phase
    shift before the final recombination pulse. Each point is an average of two
    measurements, with gray error bars representing the standard deviation. A
    detuning of $\delta_{\text{laser}} = -2 \pi \times 1050$~kHz from the hyperfine splitting
    results in oscillations which appear within an envelope whose shape is
    governed by the velocity distribution.}
\end{figure}

\begin{figure*}[t!]
  \includegraphics{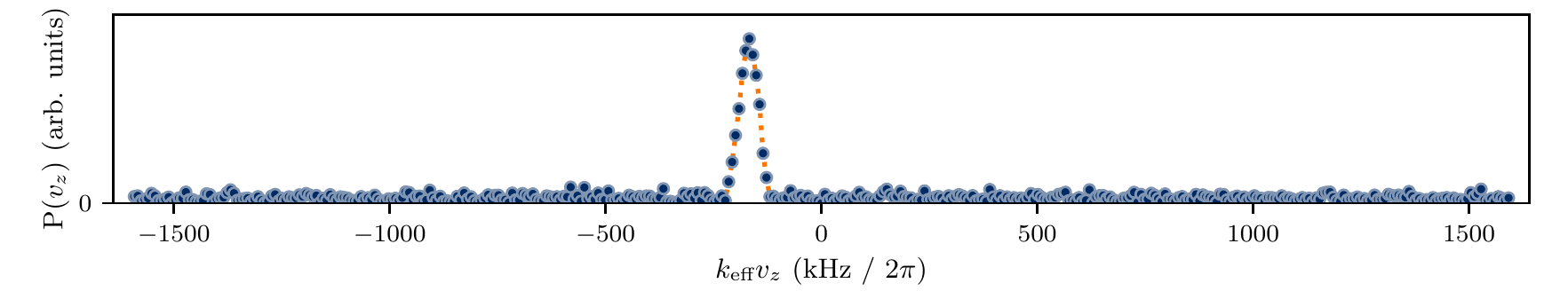}
  \caption{\label{fig:fulltrace} Absoute value of raw FFT of data in Figure
    \ref{fig:data1} after baseline subtraction, prior to subtracting the
    $\delta_{\text{laser}} = -2 \pi \times 1050$~kHz laser detuning from the
    hyperfine splitting, which manifests as a -410~mm~s$^{-1}$ shift to the
    center of the velocity distribution. The quadrature measurement is able to
    resolve the sign of the displacement, so there is no component with an
    opposite shift.}
\end{figure*}

Our experimental apparatus \cite{Dunning2015,Dunning2014} is shown schematically
in Figure \ref{fig:apparatus}. A 3D magneto-optical trap (MOT) of $^{85}$Rb
atoms is formed using a cooling laser detuned by the order of the natural
linewidth to the red of the $\transition{3}{4}$ cycling transition, and a repump
laser locked to the $\transition{2}{3}$ transition ensures that atoms are not
lost from the cooling cycle to the $\upperket{3}$ state. After a 1\,s loading
time, the magnetic field gradient is extinguished and the cooling laser reduced
in power, leaving the atoms to undergo sub-Doppler cooling in an optical
molasses. After a further period of typically 11~ms, the repump laser is
extinguished and the cooling beam optically pumps the atoms into the
$\lowerket{2}$ state for $\sim 4$~ms with a $1/e$ time constant on the order of
$100~\mu$s \cite{Note2}.  The power levels and timescales are varied
in order to achieve different sample temperatures, but the time between magnetic
field extinction and the interferometry sequence is always kept to 15~ms in
order to ensure that the magnetic field, which continues to vary slightly
through eddy effects, is correctly nulled at the time of the interferometer
sequence by additional shim coils such that the magnetic sublevels are
degenerate to within $<2\pi \times 50$~kHz.

Our interferometry is conducted with two-photon Raman transitions between the
$\lowerket{2}$ and $\lowerket{3}$ hyperfine ground states which have a frequency
splitting of $\omega_0=2\pi \times 3.036$~GHz. Two counter-propagating laser
beams, differing in frequency by ${\omega_0+\delta_{\text{laser}}}$, where
$\delta_{\text{laser}}$ is a variable detuning, are detuned from the
$\transition{3}{4}$ transition by $\Delta_{\text{1-photon}} \approx 2 \pi \times
5$~GHz. This
allows long-lived ground states to be used, while the Doppler sensitivity is
characterized by an effective wavenumber $k_{\text{eff}} =
2\pi/\lambda_{\text{eff}}$, with $\lambda_{\text{eff}} \approx(780 / 2)$~nm.

These beams are derived from an amplified free-running distributed feedback
(DFB) laser (Eagleyard EYP-DFB-0780-00080-1500-BFW01-0005), with the higher
frequency beam formed by the first upper diffracted order from a 310~MHz
acousto-optic modulator (AOM) and the lower frequency shift achieved with a
2.7~GHz electro-optic modulator (EOM), from which the carrier frequency is
suppressed by a fiber Mach-Zehnder interferometer \cite{Cooper2013}. The high
frequency EOM sideband remains and, while playing no part in the Raman
transition, contributes to the AC Stark shift.

The beams are separately amplified with tapered amplifier diodes, combined on a
further AOM for fast (${\sim 50}$~ns) shuttering, and then separated again by
polarization in order to be coupled into separate fibers which transfer the
light to the MOT chamber. At the fiber outputs the beams are orthogonally,
linearly, polarized and pass through Topag GTH-4-2.2 refractive beam shapers
which give them a 1.4~mm square profile with $\sim 15\%$ intensity variations
across the MOT cloud at the focus of 750~mm focal length lenses. This allows for
an intensity around 5~W~cm$^{-2}$ per frequency component over the
interferometry region.

The interferometer read-out is performed by illuminating the atoms with the MOT
cooling light for $300~\mu$s and collecting the fluorescence onto a Hamamatsu
H7422-50 photo-multiplier tube (PMT). The atoms are then pumped back into the
$\lowerket{3}$ for $\sim50\mu$s by the resonant repump light \cite{Note2}, and
then re-illuminated by the cooling light for a further $300~\mu$s. Exponential
decay functions are fitted to the cooling fluorescence PMT signals, and the
ratio of their amplitudes is used as a measure of the fraction of atoms
transferred to the $\lowerket{3}$ state during the interferometer. Full details
and characterisation of the read-out process can be found in
\cite{Dunning2014b}.

The AC Stark shift is determined by measuring the population transfer of a
$\pi-$pulse measured as a function of the Raman detuning $\delta_{\text{laser}}$
before interferometry. The detuning is then set to coincide with the peak
transfer. This detuning introduces a shift in the frequency domain of the signal
which should be subtracted from the derived velocity distribution in order to
extract a representative result. When this shift is greater than the width of
the Doppler profile, the DC component of the signal does not coincide with any
atomic velocity so that there is no ambiguity in analysis.

\section{Results\label{sec:results}}

\begin{figure}
  \includegraphics{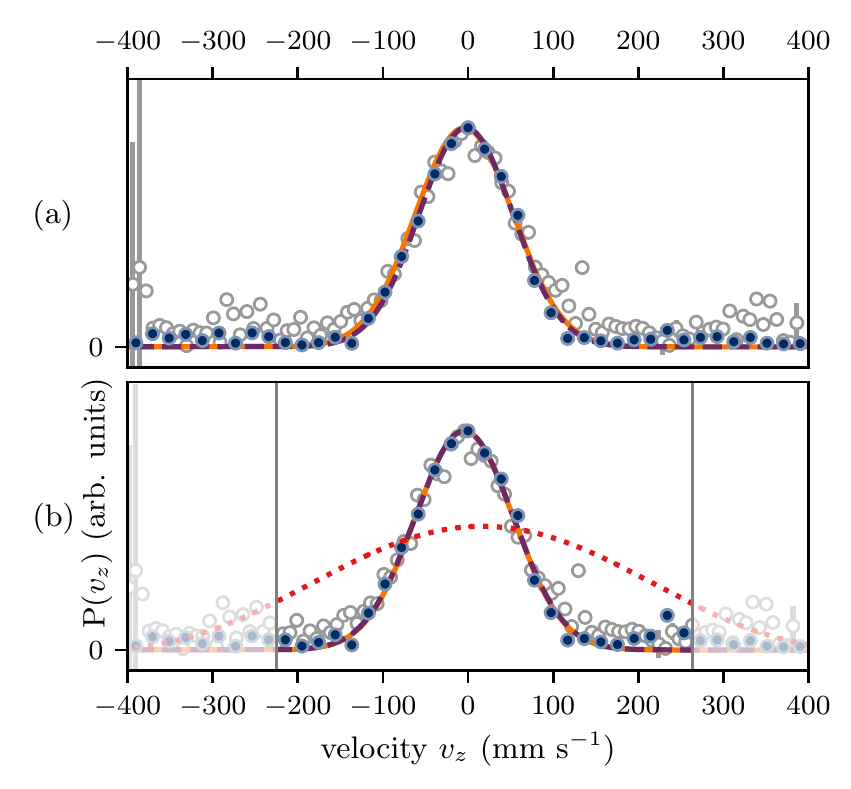}
  \caption{\label{fig:data2}Absolute value of the FFT of data in Figure
    \ref{fig:data1}, filled (blue) circles, overlaid upon a velocity profile
    determined by Raman Doppler spectroscopy, empty (gray) circles, in units of
    velocity. Solid (orange) lines are a Gaussian fit to the spectroscopic data,
    with a temperature of $32.9\pm1.4~\mu$K, and dashed (purple) lines show
    Gaussian fits to the interferometric data. Plot (a) uses raw FFT data and
    the Gaussian fit is narrower than the spectroscopic measurement with a
    temperature of $30.4\pm1.4~\mu$K. Plot (b) has a correction factor applied,
    multiplying each point by the reciprocal of the theoretical amplitude
    $\mathcal{A}(k_{\text{eff}}v_z)$ based on our experimental parameters,
    overlaid as a dashed (red) line. Only points in the $\mathcal{A}>0.4$ range
    (highlighted) were corrected to avoid amplifying noise at the extremities,
    bringing the fitted temperature to $33.4\pm1.6~\mu$K in agreement with the
    fit to the spectroscopic data. The AC Stark shift-induced offset from the
    spectroscopic data has been subtracted, centering it on the interferometric
    data at $v_z=-4.5$~mm~s$^{-1}$ to better compare their shapes.}
\end{figure}

Figure \ref{fig:data1} shows typical output from the interferometer, with a
detuning $\delta_{\text{laser}}=-2 \pi \times 1050$~kHz introduced from the
two-photon Raman resonance to offset the AC Stark shift during the pulses. The
in-phase and quadrature fringes correspond to $-\mathcal{S}_I$ and
$\mathcal{S}_Q$ from Equation (\ref{eq:combinedsignal}), with the inversion of
the in-phase component arising from the additional rotation by $\pi$. The fast
Fourier transform (FFT) of these data gives the velocity distribution, without
need for further determination of fringe visibility, and is shown, in units of
velocity, in Figure \ref{fig:fulltrace}.

The velocity resolution of the FFT is determined by the range of $\tau$, limited
in principle to the time it takes atoms to leave the interaction region. Our
data, in the range ${|\tau|<10}$~$\mu$s (spanning $\Delta\tau = 20$~$\mu$s),
give a velocity resolution of $\delta v = 1/(k_{\text{eff}}\Delta \tau) \approx
20$~mm~s$^{-1}$. The value $\delta \tau$ by which $\tau$ is incremented between
adjacent data points, in this case $\delta \tau = 50$~ns, determines the range
of velocities that can be measured $\Delta v = 1/(k_{\text{eff}} \delta t)$,
although oversampling reduces the sensitivity of the measurement to the noise on
any individual data point.

The absolute values of the FFT data \cite{Note1} are reproduced with the
two-photon detuning subtracted in Figure \ref{fig:data2}a, where they are
overlaid upon a measurement made by conventional Raman Doppler spectroscopy.
Fitted Gaussian profiles, with temperatures of $30.4\pm1.4~\mu$K and
$32.9\pm1.4~\mu$K respectively, are shown.

The FFT profile, shown in Figure \ref{fig:data2}(a), requires a small correction
to account for a slight dependence of the fringe amplitude
$\mathcal{A}(k_{\text{eff}}v_z)$ upon atomic velocity. Figure \ref{fig:data2}(b)
shows this velocity dependence (dotted line), together with the corrected
velocity distribution which yields the same temperature as the Doppler
measurements, with notably enhanced signal-to-noise ratio. This velocity
dependence, together with some parasitic contributions to the interferometer
output at large detunings, are discussed in Appendix
\ref{sec:analyt-expr-interf}, and impose an effective upper limit on the range
of velocities that can be measured. As these effects depend on the Rabi
frequency $\Omega_R$, higher intensity Raman beams could be used to extend this
limit. The profile in this instance is slightly displaced to account for a
difference of $2 \pi \times 40$~kHz between the assumed laser detuning
$\delta_{\text{laser}}$ and the AC Stark shift recorded for this experiment.

The Gaussian fit to the corrected data is centered at $v_z=-4.5 \pm
1.3$~mm~s$^{-1}$. This is in agreement with a second measurement, shown in
Figure \ref{fig:FFT2}, centered at $v_z = -5.1 \pm 0.7$~mm~s$^{-1}$. We note
that the two-photon recoil velocity for $^{85}$Rb is 12~mm~s$^{-1}$; and there
could be an impulse imparted to the cloud as the magnetic field is terminated.

\section{Discussion\label{sec:discussion}}

\begin{figure}[b]
  \includegraphics{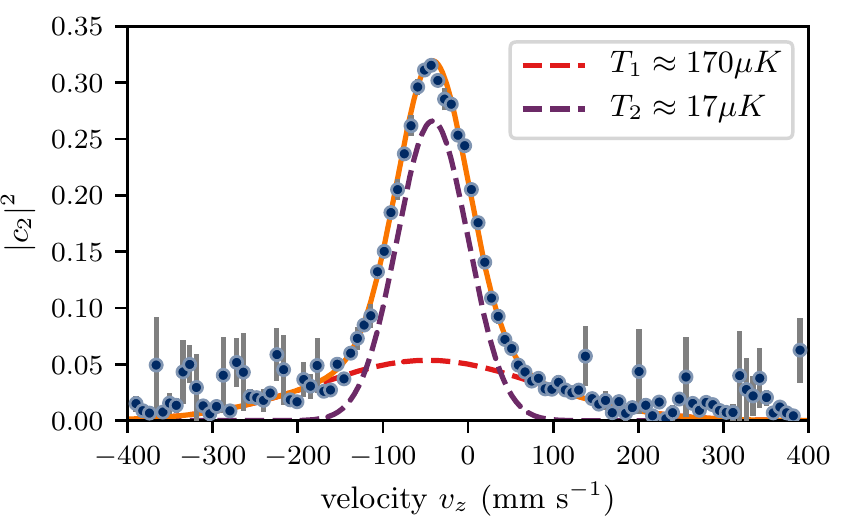}
  \caption{\label{fig:raman-veloc}Doppler spectroscopy velocity profile taken
    with Raman $\pi$-pulses with a single-photon detuning $\sim 2 \pi \times
    7$~GHz and a Rabi frequency $\Omega_R \approx 2\pi \times 50$~kHz. A single
    Gaussian does not make a good fit, but a two Gaussian distribution fits
    well. The solid (orange) line is the sum of the two dashed Gaussians, a cold
    central peak and a broader background which we have previously attributed to
    inhomogenous sub-Doppler cooling. The resonance is AC Stark shifted,
    resulting in the large offset of the distribution from $v_z=0$.}
\end{figure}

The velocity distribution determined from our interferometric measurements
agrees well for a warm sample with that obtained by conventional Doppler
spectroscopy using low power Raman pulses (single-photon detuning $\sim 2 \pi
\times 15$~GHz, Rabi frequency $\Omega_R \approx 2 \pi \times 25$~kHz). Whereas
the interferometric measurements are well represented by a single Gaussian,
however, the Doppler measurements show an additional, broader, component. This
is particularly noticeable when a cooler sample, such as that shown in Figure
\ref{fig:raman-veloc}, is probed with a strong Raman field (single-photon
detuning $\sim 2\pi \times 7$~GHz, Rabi frequency $\Omega_R \approx 2 \pi \times 50$~kHz),
yielding a central Gaussian distribution with a temperature of $17.5\pm0.9~\mu$K
superimposed upon a much broader background.

\begin{figure}[t]
  \includegraphics{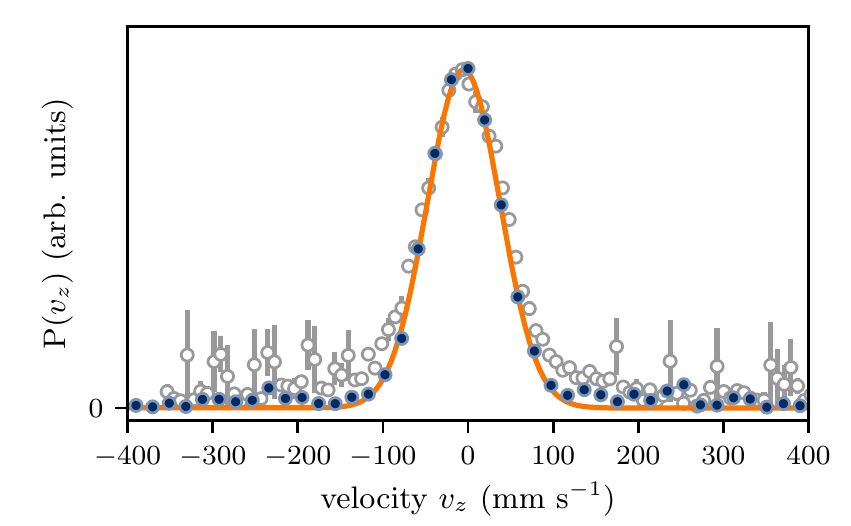}
  \caption{\label{fig:FFT2} Interferometric velocimetry measurement, filled
    circles (blue), overlaid upon the Doppler spectroscopy profile from Figure
    \ref{fig:raman-veloc}, empty circles (gray). The measurements were taken
    under the same conditions, with the offset subtracted from the spectroscopic
    data to center them on the interferometric profile at
    $v_z=-5.1$~mm~s$^{-1}$. The solid (orange) line shows a Gaussian fit to the
    interferometric data, with a temperature of $18.3\pm0.6~\mu$K corresponding
    closely to the $17.5\pm0.9~\mu$K colder Gaussian fitted in Figure
    \ref{fig:raman-veloc}. The signal-to-noise ratio of the interferometric
    measurement is good, and does not show signs of the broad background evident
    in the spectroscopic measurement.}
\end{figure}

Interferometric measurement under the same conditions, shown in Figure
\ref{fig:FFT2}, does not display this broad component, but modeling suggests
that this is not a limitation of the interferometric technique. We have
previously attributed the broad background to inhomogeneous sub-Doppler cooling
\cite{Townsend1995}; such a distribution might also result if the Doppler
technique detected warmer, untrapped atoms outside the region interrogated by
the interferometer. The dependence upon the strength of the probe laser in the
Doppler measurements, however, suggests that the broadening is an artefact of
conventional Doppler methods, perhaps due to off-resonant excitation
\cite{Hughes2017}. This is consistent with several determinations of the atom
cloud temperature from measurements of the coherence length of the atomic
wavepacket by measuring the fringe contrast as a function of wavepacket
separation, in each case yielding a temperature below that estimated by Doppler
\cite{Parazzoli2012} or time-of-flight
\cite{Saubamea1997,Featonby1998,Marechal2000} methods.

Time-of-flight measurements are often used for colder samples and condensates
\cite{Anderson1995}, but these are limited by the physical extent of the cloud
and the imaging resolution \cite{Brzozowski2002}. In practice this means that
the time of expansion required to measure the coldest distributions is typically
$\sim10$~ms, limiting its usefulness when studying dynamic behaviour such as in
\cite{Afek2017}. Both interferometric and Doppler measurements can be performed
faster; such measurements with a Fourier transform limited resolution equivalent
to Figure \ref{fig:FFT2} for a 10nK cloud could be made in as little as
$500~\mu$s. However, the Doppler measurement requires a continuous interaction
for this time, increasing the probability of the artefacts we have observed
while limiting the resonance to a small number of atoms and reducing the
signal-to-noise ratio.

Interferometric measurement, in contrast, ideally involves interactions that
last for a small fraction of the total measurement time and interact uniformly
with the entire velocity and spatial distribution of the atom cloud so that, on
average, half of the atoms contribute to the signal, limited by the finite range
that can be addressed in practice. Interferometric velocimetry is hence a
particularly effective complement to existing methods and is particularly
suitable for colder atom samples in which artefacts such as off-resonant
excitation, saturation and scattering force heating would otherwise distort the
measured velocity distributions. It uses techniques, apparatus and, in some
cases \cite{Deh2009, Featonby1998}, datasets that are often already to hand.

\section{Conclusion\label{sec:conclusions}}

We have described the use of Ramsey matterwave interferometry for the
measurement of the velocity distribution, and hence translational temperature,
of ultracold rubidium atoms. By using an asymmetrical 3-pulse arrangement with
switchable pulse phases, we record quadrature signals over both positive and
negative effective interferometer durations. The Fourier transform, with
correction for the residual Doppler effect within the interferometer pulses
themselves, then reveals the atomic velocity distributions with good fidelity
because the whole atomic sample contributes to each data point. The technique
can distinguish between positive and negative velocities with a quadrature
measurement, and is more effective at lower cloud temperatures. It is in many
ways complementary to conventional techniques of Doppler-sensitive spectroscopy
and time-of-flight measurement, as it is not limited by off-resonant excitation
\cite{Hughes2017} or the physical extent of the atom cloud and measurement beam
\cite{Brzozowski2002}.

\appendix
\section{Analytical model for interferometer
  output\label{sec:analyt-expr-interf}}

Here we derive analytical expressions for the output of a Mach-Zehnder atom
interferometer, with a temporal asymmetry $\tau$ between the two periods of free
evolution. We consider two ways of introducing this asymmetry:

\begin{enumerate}[label=(\alph*)]
\item keeping the total interferometer time constant such that $T_1 = T -
  \tau/2$ and $T_2 = T + \tau / 2$, illustrated by the dashed (blue) line in
  Figure \ref{fig:sequences}c, and
\item keeping one separation constant such that $T_1 = T$ and $T_2 = T + \tau$,
  illustrated by the dotted (red) line in Figure \ref{fig:sequences}c.
\end{enumerate}

We model the effect of these sequences on a two-level atom with state amplitudes
$c_1$ and $c_2$, assuming completely coherent evolution and treating pulses in
the rotating wave approximation \cite{Rabi1954}.

Between the pulses, the superposition accrues phase at a rate equal to the
atom--laser detuning $\Delta = \delta_{\text{laser}} + \delta_{\text{doppler}}$,
in which we include the laser detuning from the state splitting
$\delta_{\text{laser}}$ in addition to the velocity-dependent detuning
$\delta_{\text{doppler}} = k v_z$. Free evolution for a period $T$ can then be
represented by the matrix

\begin{equation}
  \label{eq:free-evolution}
  \mathbf{U}(T)=
  \left( 
    \begin{array}{cc}
      e^{-i \Delta \cdot T / 2} & 0 \\ 0 & e^{i \Delta \cdot T / 2}
    \end{array}
  \right),
\end{equation}

\noindent acting on a state vector $\left(\begin{array}{cc}c_1 \\
    c_2 \end{array}\right)$.

During the pulses, the superposition undergoes a rotation in Hilbert space
\cite{Feynman1957} whose rate and orientation are determined by the on-resonance
Rabi frequency $\Omega$ and the detuning $\Delta_{\text{ac}} = \Delta -
\delta_{\text{ac}}$, which differs from the inter-pulse detuning by a term
$\delta_{\text{ac}}$ due to the AC Stark shift.

The effect of a pulse of duration $T$ can be solved analytically and, following
the formalism of Stoner et al \cite{Stoner2011}, can be represented by the
matrix

\begin{equation}
  \label{eq:stoner-operator}
  \mathbf{\Omega}(T,\phi_L) = \left(
    \begin{array}{cc}
      C(T) & -i S(T, \phi_L) \\
      -i S^*(T, \phi_L) & C^*(T) \\
    \end{array}
  \right)
\end{equation}

\noindent to act on a state vector, with

\begin{subequations}
  \begin{align}
    \label{eq:matrix-components}
    C(T) &= \cos \left(\frac{T}{2} \sqrt{\Delta_{\text{ac}} ^2+\Omega ^2} \right) \nonumber\\
         & \quad+  i \frac{\Delta_{\text{ac}} }{\sqrt{\Delta_{\text{ac}}^2+\Omega^2}}\sin \left(\frac{T}{2} \sqrt{\Delta_{\text{ac}} ^2+\Omega ^2} \right),\\
    S(T, \phi_L) &= \frac{\Omega e^{i \phi _L}}{\sqrt{\Delta_{\text{ac}}^2+\Omega^2}}\sin \left(\frac{T}{2} \sqrt{\Delta_{\text{ac}} ^2+\Omega ^2} \right).
  \end{align}
\end{subequations}

\noindent Here we have included an explicit dependence on the laser phase
$\phi_L$.

After a pure state ($c_1 = 1$, $c_2 = 0$) has been subjected to an
interferometer sequence, the excited state probability $|c_2|^2$ can then be
calculated with matrix multiplication.

We denote the $n$th pulse with subscripts $\mathbf{\Omega}_n, C_n, S_n$ etc. so
that the output of a 3-pulse interferometer with pulse separations $T_{1,2}$ is

\begin{equation}
  \label{eq:3-pulse-output}
  \begin{split}
    |c_2|^2 &= \biggl|\left(\begin{array}{cc}0 & 0 \\ 0 & 1 \end{array}\right)
    \mathbf{\Omega}_3\mathbf{U}\left(T_2\right)\mathbf{\Omega}_2
    \mathbf{U}\left(T_1\right)\mathbf{\Omega}_1\left(\begin{array}{cc}1 \\ 0 \end{array}\right)\biggr|^2,\\
    &= |S_1|^2|S_2|^2|S_3|^2 + |C_1|^2|S_2|^2|C_3|^2 \\
    &\quad- 2 \Re\left[ e^{i \Delta (T_2 - T_1)} C_1 S_1 \left( S_2^* \right)^2 C_3^* S_3 \right]\\
    &\quad+|S_1|^2|C_2|^2|C_3|^2 + |C_1|^2|C_2|^2|S_3|^2 \\
    &\quad+ 2 \Re\left[ e^{i \Delta (T_1 + T_2)} C_1^* S_1^* \left( C_2^* \right)^2 C_3^* S_3 \right] \\
    &\quad+ 2 \left( |C_3|^2 - |S_3|^2 \right)\Re\left[ e^{- i \Delta \cdot T_1} C_1 S_1 C_2 S_2^* \right] \\
    &\quad+ 2 \left( |C_1|^2 - |S_1|^2 \right) \Re\left[ e^{i \Delta \cdot T_2}
      C_2^* S_2^* C_3^* S_3 \right].
  \end{split}
\end{equation}

To proceed we note that, for the Mach-Zehnder interferometers under
consideration,

\begin{subequations}
  \begin{align}
    C_3 &= C_1,\label{eq:conditions1}\\
    S_3 &= e^{i\phi}S_1,\label{eq:conditions2}\\
    \arg(S_1) &= \arg(S_2)\nonumber\label{eq:conditions3} \\
    \Rightarrow S_1^* S_2 &= S_1 S_2^* = |S_1||S_2|,
  \end{align}
\end{subequations}

\noindent where $\phi$ is an advance in the laser phase introduced prior to the
final pulse. This allows us to write Equation (\ref{eq:3-pulse-output}) as

\begin{equation}
  \label{eq:3-pulse-output-simplified}
  \begin{aligned}
    |c_2|^2 &= |S_1|^4|S_2|^2 + |C_1|^4|S_2|^2 + 2 |S_1|^2|C_2|^2|C_1|^2\\
    &- 2 |C_1|^2|S_1|^2|S_2|^2\cos[ \Delta (T_2 - T_1) + \phi ]\\
    &+ 2 |S_1|^2\Re[C_1^2C_2^2] \cos[\Delta (T_1 + T_2) + \phi ]\\
    &+ 2 |S_1|^2\Im[C_1^2C_2^2] \sin[\Delta (T_1 + T_2) + \phi ] \\
    &+ 2 ( |C_1|^2 - |S_1|^2 )|S_1||S_2|\Re[C_1 C_2]\\
    &\quad\quad\quad\quad\quad\times[\cos(\Delta \cdot T_1) + \cos(\Delta \cdot T_2 + \phi)] \\
    &+ 2 ( |C_1|^2 - |S_1|^2 )|S_1||S_2|\Im[C_1 C_2]\\
    &\quad\quad\quad\quad\quad\times[ \sin(\Delta \cdot T_1) +\sin(\Delta \cdot
    T_2 + \phi) ].
  \end{aligned}
\end{equation}

Let us first consider case (a), the interferometer keeping the free-evolution
time constant, such that $T_{1,2} = T \mp \tau/2$. Noting that

\begin{subequations}
  \begin{align}
    \label{eq:trig-relations}
    &\sin\left[ \Delta\left( \frac{\tau}{2} + T \right) + \phi \right] + \sin \left[ \Delta\left( T - \frac{\tau}{2} \right) \right] =\nonumber \\
    &\quad\quad 2 \sin \left( \Delta \cdot T + \frac{\phi}{2} \right) \cos \left( \Delta \frac{\tau}{2} + \frac{\phi}{2} \right),\text{ and}\\
    &\cos\left[ \Delta\left( \frac{\tau}{2} + T \right) + \phi \right] + \cos \left[ \Delta\left( T - \frac{\tau}{2} \right) \right] =\nonumber \\
    &\quad\quad 2 \cos \left( \Delta \cdot T + \frac{\phi}{2} \right) \cos \left( \Delta \frac{\tau}{2} + \frac{\phi}{2} \right), 
  \end{align}
\end{subequations}

the output can be shown to be

\begin{align}
  \label{eq:output1b}
  |c_2|^2 &= \frac{1}{2}\bigl\{\mathcal{C} - \mathcal{A} \cos\left(\Delta \tau + \phi\right) \nonumber\\
          &\quad\quad\quad + \mathcal{B}\cos\left[\frac{1}{2}\left(\Delta \tau + \phi\right)\right]  \bigr\}.
\end{align}

with

\begin{subequations}
  \begin{align}
    \label{eq:fixed-time-abc}
    \mathcal{A} &= 4|C_1|^2|S_1|^2|S_2|^2,\\
    \mathcal{B} &= 8 ( |C_1|^2 - |S_1|^2 )|S_1||S_2|\nonumber \\
                &\quad\times\bigl[\Re(C_1 C_2)\cos(\Delta \cdot T + \phi/2)\nonumber \\
                &\quad\quad + \Im(C_1 C_2)\sin(\Delta \cdot T + \phi/2)\bigr],\\
    \mathcal{C}  &= 4 |C_1|^2|C_2|^2|S_1|^2 + 2 |C_1|^4|S_2|^2 + 2 |S_1|^4|S_2|^2 \nonumber \\
                &\quad + 4 |S_1|^2 \bigl[ \Re(C_1^2C_2^2)\cos(\Delta \cdot T + \phi) \nonumber \\
                &\quad\quad + \Im(C_1^2C_2^2)\sin(\Delta \cdot T + \phi) \bigr].
  \end{align}
\end{subequations}

The output has sinusoidal components in $\tau$ with velocity-dependent frequency
$\Delta$, and amplitude $\mathcal{A}(\Omega, \Delta_{\text{ac}})$, purely
constructed from elements of the pulse matrices and thus only dependent on the
offset from the light-shifted resonance (with the overall scale determined by
the on-resonance Rabi frequency). There is no detuning-dependent phase shift to
these harmonic components, though parasitic subharmonics of amplitude
$\mathcal{B}(\Omega, \Delta, \Delta_{\text{ac}}, \phi, T)$ become significant at
large detunings.

\begin{figure}
  \includegraphics{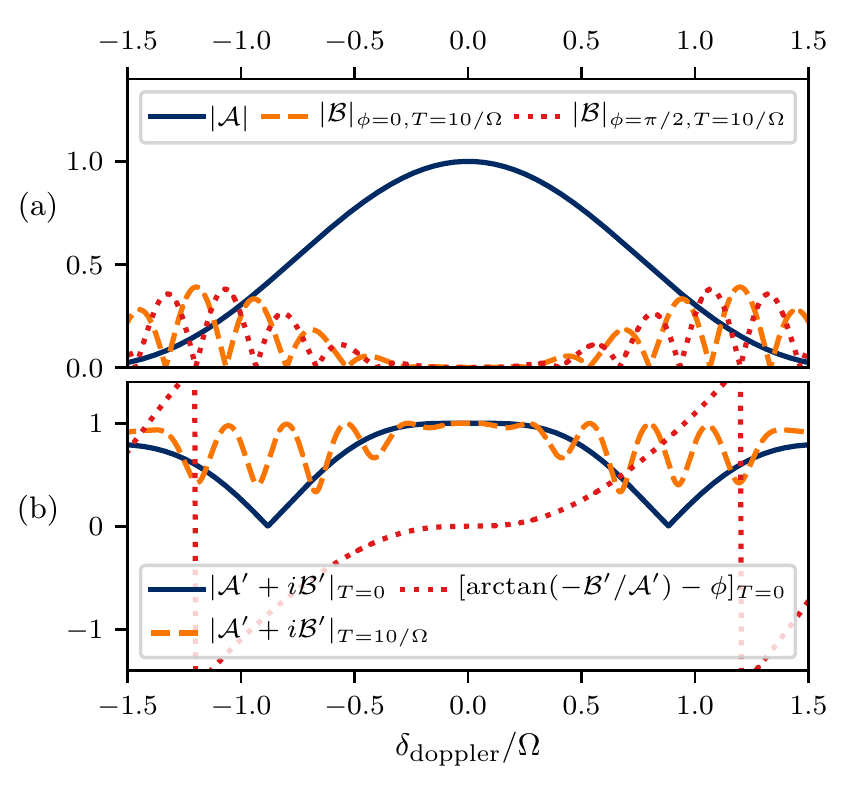}
  \caption{\label{fig:amplitudes}Amplitudes of frequency components in
    interferometer outputs as a function of Doppler detuning, in units of the
    on-resonance Rabi frequency. In these plots $\delta_{\text{ac}} =
    \delta_{\text{laser}}$; when this is not the case, the profile is shifted in
    position by the difference. (a) Amplitudes for $T_1=T-\tau/2$,
    $T_2=T+\tau/2$ interferometer. Solid (blue) line shows amplitude of
    $\delta_{\text{doppler}}$ harmonics. Dashed (orange) and dotted (red) lines
    show the amplitude of $\delta_{\text{doppler}}/2$ subharmonics with $\phi=0$
    and $\phi=\pi/2$ respectively, when $T=10/\Omega$. (b) Amplitudes for $T_1 =
    T$, $T_2=T+\tau$ interferometer. No subharmonics are present, but the
    $\delta_{\text{doppler}}$ harmonics receive a residual detuning-dependent
    phase shift, plotted as a dotted (red) line for $T=0$. Solid (blue) line
    shows the amplitude profile when $T=0$. Dashed (orange) line shows the
    profile for $T=10/\Omega$, demonstrating the appearance of an amplitude
    modulation in lieu of the presence of subharmonics.}
\end{figure}

With $\mathcal{A} = \mathcal{C} = 1$, $\mathcal{B} = 0$, and
$\delta_{\text{laser}} = 0$, Equation (\ref{eq:output1b}) resembles the
analogous Ramsey output in the integrand of Equation (\ref{eq:inphaseoutput}),
albeit with an inversion arising from the additional rotation by $\pi$. As long
as $\mathcal{A} \gg \mathcal{B}$ then, as in Equation (\ref{eq:inphaseoutput}),
the atomic velocity distribution is well mapped onto the frequency domain and
the scaling by $\mathcal{A}$ can be corrected for by multiplying through by its
reciprocal.
 
The magnitudes of $\mathcal{A}$ and $\mathcal{B}$ are plotted as functions of
$\delta_{\text{doppler}}/\Omega$ in Figure \ref{fig:amplitudes}a which shows
that this criterion is satisfied for $-0.4 < \delta_{\text{doppler}}/\Omega <
0.4$, where $|\mathcal{B}/\mathcal{A}|<0.1$. Doppler profiles falling within
this window will thus incur little distortion from subharmonics, which would act
to artificially narrow broader distributions. As the interferometer time is made
longer by increasing $T$, the oscillations in $\mathcal{B}$ become more rapid
but remain within the same envelope.

It should be noted that the velocity distribution will be centered about
$\delta_{\text{laser}}$ in the frequency domain, so the ambiguity of the DC
component $\mathcal{C}(\Omega, \Delta, \Delta_{\text{ac}},\phi,T)$ can be
negated by setting $\delta_{\text{laser}}$ much larger than the width of the
Doppler profile and subtracting it off in analysis.

\begin{figure}[t]
  \includegraphics{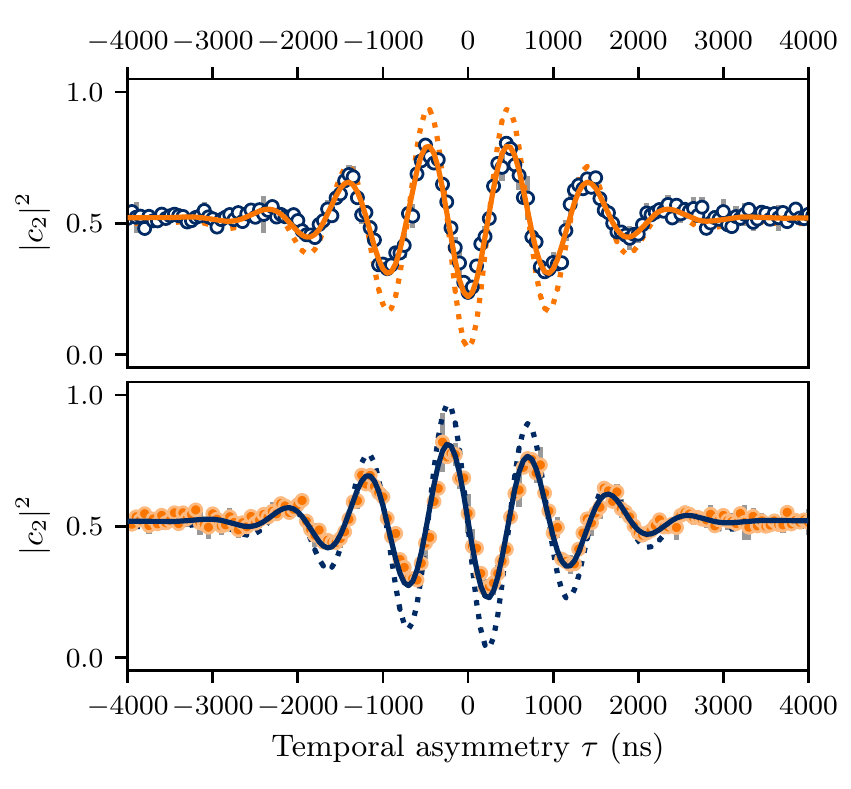}
  \caption{\label{fig:theory-verif}Empty (blue) and filled (orange) circles
    respectively show the in-phase and quadrature data from Figure
    \ref{fig:data1}. Dashed lines show the output from Equation
    (\ref{eq:output1b}) integrated over a 33$\mu$K velocity distribution, as per
    Figure \ref{fig:data2}, with $\delta_{\text{ac}}=-2 \pi \times 1010$ kHz,
    $\delta_{\text{laser}}=-2 \pi \times 1050$ kHz, and $\Omega=2 \pi \times
    685$ kHz taken from experimental parameters. Solid lines are the same model
    vertically centered on the experimental data and scaled by a factor of 0.63.
    The contrast loss is expected, as the model does not account for sources of
    dephasing and finite interaction region present in the real experiment, but
    the shape otherwise agrees well. We attribute the slight vertical offset in
    the data to incoherent excitation or imperfect state preparation.}
\end{figure}

Equation (\ref{eq:output1b}) can be evaluated as a function of $\tau$ and
integrated over a distribution of velocities (and hence detunings $\Delta$) to
predict the interferometer output for the given velocity distribution. This is
shown by the dotted lines in Figure \ref{fig:theory-verif}, which are calculated
from the distribution measured in section \ref{sec:results} with the same
experimental parameters at $\phi=0$ and $\phi=\pi/2$.

This analysis assumes completely coherent evolution, no phase or amplitude noise
on the laser, and no additional phase terms from external fields. Further, it
assumes a perfect 2-level atom, as opposed to the $^{85}$Rb Raman system with
different coupling strengths for magnetic sublevels. Empirically we observe that
these factors result in a loss of contrast but do not noticeably affect the
shape of the signal, as demonstrated by the solid lines in Figure
\ref{fig:theory-verif} which have been vertically centered on the experimental
data from Figure \ref{fig:data1} and scaled by a factor of 0.63, demonstrating
good agreement. We attribute the slight positive offset of the experimental data
to imperfect state preparation or incoherent (single-photon) excitation.

Let us now treat case (b), introduced at the beginning of this section as the
interferometer in which $T_1 = T$ is kept constant and $T_2 = T+\tau$, in a
similar manner. By expanding the trigonometric functions of Equation
(\ref{eq:3-pulse-output-simplified}), we can isolate the $\tau$ dependence into
terms of $\sin(\Delta \tau + \phi)$ and $\cos(\Delta \tau + \phi)$ such that the
interferometer output can be written in the form

\begin{equation}
  \label{eq:output2b}
  |c_2|^2 = \frac{1}{2}\bigl\{\mathcal{C}^\prime + \mathcal{A}^\prime \cos\left(\Delta \tau + \phi\right)
  + \mathcal{B}^\prime\sin\left(\Delta \tau + \phi\right)  \bigr\},
\end{equation}

with

\begin{subequations}
  \begin{align}
    \label{eq:varying-time-abc}
    \mathcal{A}^\prime &= - 4|C_1|^2|S_2|^2|S_1|^2\nonumber \\
                       &\quad + 4|S_1|^2 \bigl[\Im(C_1^2 C_2^2) - |S_1|| S_2|\Im(C_1 C_2) \bigr] \sin (\Delta \cdot T)\nonumber \\
                       &\quad + 4|S_1| ^2 \bigl[\Re(C_1^2 C_2^2) - |S_1|| S_2|\Re(C_1 C_2) \bigr] \cos(\Delta \cdot T)\nonumber \\
                       &\quad + 4|C_1|^2|S_1|| S_2|\bigl[\Im(C_1 C_2)\sin (\Delta \cdot T)\nonumber \\
                       &\quad\makebox[\widthof{$+ 2|C_1|^2|S_1|| S_2|\bigl[$}]{}\quad + \Re(C_1 C_2)\cos(\Delta \cdot T) \bigr],\\
    \mathcal{B}^\prime &=4\bigl[|S_1|^2 \Im(C_1^2 C_2^2) + (|C_1|^2 - |S_1|^2)\nonumber\\
                       &\makebox[\widthof{$=2\bigl[$}]{}\quad \times |S_1||S_2|\Im(C_1 C_2) \bigr] \cos(\Delta \cdot T)\nonumber\\
                       &\quad - 4\bigl[|S_1|^2 \Re(C_1^2 C_2^2) + (|C_1|^2 - |S_1|^2)\nonumber\\
                       &\quad\makebox[\widthof{$- 2\bigl[$}]{}\quad \times |S_1||S_2|\Re(C_1 C_2) \bigr] \sin(\Delta \cdot T),\\
    \mathcal{C}^\prime &= 2|C_1|^4|S_2|^2 + 2|S_1|^4|S_2|^2 + 4|C_1|^2|C_2|^2|S_1|^2 \nonumber\\
                       &\quad - 4|S_1|^3|S_2|\bigl[\Re(C_1 C_2) \cos(\Delta \cdot T)\nonumber\\
                       &\quad\makebox[\widthof{$- 4|S_1|^3|S_2|\bigl[$}]{}\quad +\Im(C_1 C_2)\sin(\Delta \cdot T) \bigr]\nonumber \\
                       &\quad + 4|C_1|^2|S_1||S_2|\bigl[\Re(C_1 C_2)\cos(\Delta \cdot T) \nonumber\\
                       &\quad\makebox[\widthof{$ + 4|C_1|^2|S_1||S_2|\bigl[$}]{}\quad + \Im(C_1 C_2)\sin (\Delta \cdot T) \bigr].
  \end{align}
\end{subequations}

Subharmonic components are no longer present in this output, but quadrature
terms are, introducing an effective detuning-dependent phase shift
$\arctan\left( -\mathcal{B}^\prime/\mathcal{A}^\prime \right)$ which, unlike the
Ramsey interferometer \cite{Carey2017}, has a flat gradient through
$\Delta_{\text{ac}} = 0$. $|\mathcal{A}^\prime+i\mathcal{B}^\prime|$ gives the
amplitude of the harmonic components which, for $T>0$, exhibits an oscillatory
modulation, the envelope of which resembles that of the subharmonics present in
case (a).

The case of $T=0$ warrants special attention. This reduces to a 2-pulse
$3\pi/2$--$\pi/2$ interferometer, restricting the pulse separation to $\tau>0$.
This can still yield good velocimetry results when the gradient of the
detuning-dependent phase shift is small, as it is about $\Delta_{\text{ac}}=0$,
unlike in a Ramsey interferometer \cite{Carey2017}. This case is considered in
Appendix \ref{sec:enhanc-rams-interf}.

\section{Enhanced Ramsey interferometer\label{sec:enhanc-rams-interf}}

In a two-pulse interferometer, data collection is inherently restricted to the
$\tau>0$ window, mathematically expressed by multiplying the time domain output
by the Heaviside step function $\Theta(\tau)$. In the frequency (velocity)
domain, this manifests as a convolution with the Fourier transform of the step
function

\begin{equation}
  \label{eq:heaviside}
  \tilde{\Theta}(\omega) = \frac{i}{\sqrt{2 \pi}\omega} + \sqrt{\frac{\pi}{2}}\delta(\omega),
\end{equation}

\noindent $\delta(\omega)$ being the Dirac delta function. This introduced an
imaginary component to the output, but leaves the velocity distribution
unaltered in the real part.

In a realistic Ramsey interferometer (Figure \ref{fig:sequences}b) where the
pulses are of finite length, setting the separation to $\tau=0$ is not
equivalent to setting the effective interferometer period to 0. The result of
this is a multiplication by $\Theta(\tau-\delta\tau)$, where $\delta\tau$ is an
offset determined by the length of the pulses. This introduces a phase factor
into the first term of Equation (\ref{eq:heaviside}), irreversibly mixing the
real part of the $\delta\tau=0$ output with the imaginary part so that the
velocity distribution is irretrievable from a single measurement
\cite{Carey2017}.

\begin{figure}
  \includegraphics{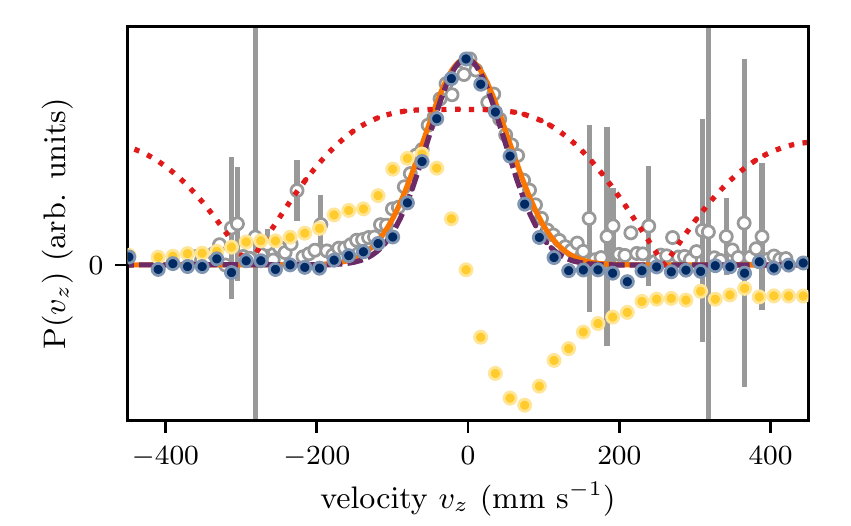}
  \caption{\label{fig:real-imag} Interferometric velocimetry measurement with
    enhanced Ramsey interferometer. Real and imaginary parts of the FFT are
    shown by dark (blue) and light (yellow) filled circles respectively,
    overlaid upon a Doppler spectroscopy profile, empty circles (gray), taken
    under the same conditions. The real part of the interferometric measurement
    shows good agreement with the spectroscopic one. The solid (orange) and
    dashed (purple) lines show Gaussian fits to the spectroscopic and
    interferometric data respectively, with corresponding temperatures of
    26~$\mu$K and 33~$\mu$K. The theoretical amplitude $|\mathcal{A}^\prime +
    i\mathcal{B}^\prime|$ from Figure \ref{fig:amplitudes}b is shown as an
    additional dashed (red) line, plotted from measured experimental
    parameters.}
\end{figure}

As touched upon at the end of Appendix \ref{sec:analyt-expr-interf}, the
asymmetric Mach-Zehnder interferometer case (b) with $T=0$ is a two-pulse
$3\pi/2$--$\pi/2$ interferometer. This has the properties of a Ramsey-type
interferometer, but with a degree of cancellation of the phase picked up during
the pulses. While the Ramsey interferometer has a linear dependence of the phase
shift on detuning, equivalent to an offset of the effective time origin, for the
$3\pi/2$--$\pi/2$ interferometer the lowest order term in the phase shift is
cubic. This gives it a flat gradient about zero detuning, a property shared by
the shift in amplitude (Figure \ref{fig:amplitudes}b).

Preliminary results, shown in Figure \ref{fig:real-imag}, show that the real
parts of the Fourier transformed interferometer provide an effective measure of
the velocity profile. Distortion due to the time domain truncation is primarily
restricted to the imaginary parts. The detuning-dependent amplitude from Figure
\ref{fig:amplitudes}b, plotted from measured experimental parameters, is shown
as a red (dotted) line; though the data have not been corrected by multiplying
by its reciprocal. The reduction in amplitude is $<10\%$ across 2 standard
deviations ($95\%$) of the Gaussian distribution fitted to the real data, which
are negligibly affected by correction (although the imaginary data are
distorted). This broad, flat, profile offers a potential advantage of what we
term the enhanced Ramsey method.

The measurement is overlaid upon a spectroscopic measurement taken under the
same conditions and they are is good agreement, with the interferometric
measurement appearing slightly narrower. We can attribute this discrepancy to
artefacts of the Doppler spectroscopy discussed in Section \ref{sec:discussion}
and sensitivity of the interferometric measurement to pulse rise-time at small
$\tau$.

\begin{acknowledgments}
  This work was supported by the EPSRC through the UK Quantum Technology Hub for
  Sensors \& Metrology under grant EP/M013294/1, and by Dstl under contracts
  DSTLX-1000091758 and DSTLX-1000097855. The data presented in this paper are
  available for download from {\bf http://doi.org/10.5258/SOTON/D0383}.
\end{acknowledgments}

\bibliography{velocimetry}

\end{document}